# Proposal for an X-Ray Free Electron Laser Oscillator with Intermediate Energy Electron Beam

Jinhua Dai    Haixiao Deng*    Zhimin Dai

*Shanghai Institute of Applied Physics, Chinese Academy of Sciences, Shanghai, 201800, P. R. China*

Harmonic lasing of low-gain free electron laser oscillators has been experimentally demonstrated in the terahertz and infrared regions. Recently, the low-gain oscillator has been reconsidered as a promising candidate for hard x-ray free electron lasers, through the use of high reflectivity, high resolution x-ray crystals. In this letter, it is proposed to utilize a crystal-based cavity resonant at a higher harmonic of the undulator radiation, together with phase shifting, to enable harmonic lasing of the x-ray free electron laser oscillator, and hence allow the generation of hard x-ray radiation at a reduced electron beam energy. Results show that fully coherent free electron laser radiation with megawatt peak power, in the spectral region of 10-25keV, can be generated with a 3.5GeV electron beam.

    The development of advanced particle accelerator technology, especially synchrotron radiation (SR) light source and free electron laser (FEL) is being made to satisfy the strongly growing demand for photon sources within material and biological sciences. The fundamental process of SR and FEL sources usually involves a relativistic electron beam passing through a transverse periodic magnetic field, e. g. the undulator, and generating transversely coherent electromagnetic radiation ranging from the infrared to hard x-ray regions, depending on the electron beam energy and the undulator period and strength. The typical undulator period of presently existing light sources is on the order of centimeters, and is limited by the practical difficulties of placing small magnets together in an alternating array and obtaining magnetic strength strong enough for coupling between the electron and the radiation. This implies that to achieve short-wavelength radiation, i.e. hard x-ray, the SR and FEL sources require a high energy electron beam, meaning large machines with substantial cost, time and effort requirement. However, the undulator also supports higher harmonics, with the radiation wavelength scaling inversely with the harmonic number. Therefore, high harmonics could be an alternative way to reach short wavelength instead of high energy electron beam. The development of intermediate energy SR light sources has been attributed to high harmonics of the undulator. Compared with the expensive large-scale SR light sources, such as European synchrotron radiation facility (ESRF), advanced photon source (APS) and super photon ring at Japan (Spring-8) where the high brightness, hard x-ray radiation is mainly generated by undulators operating at the first harmonic, the intermediate energy SR light sources operating at high harmonics have lower construction and operation costs, and are able to provide comparable performance in the 10-25keV photon energy range [1].

    In recent decades, there has been an increasing interest in the development of short-wavelength FELs, known as the 4[th] generation light source, capable of enabling new areas of science. Currently, the Linac Coherent Light Source (LCLS) [2], the word's first hard x-ray FEL is in operation as a user facility, and another four hard x-ray FEL user facilities are planned to operate in future years [3-6]. All of the cited hard x-ray FELs use self-amplified spontaneous emission (SASE) [7] as the lasing mode, which starts from the initial shot noise of the electron beam, and results in radiation with excellent spatial coherence, but with rather poor temporal coherence. Therefore, various seeded FELs [8-10] have been proposed to produce fully coherent radiation pulses approaching the hard x-ray region. It is widely believed that the high-gain single-pass FEL is the leading candidate in the pursuit of hard x-ray FEL sources. Harmonic radiation also exists in the emission spectrum of the high-gain FEL, and has been theoretically studied [11] and experimentally measured [12, 13]. However, the harmonics of the high-gain FEL are nonlinear and weak. The output power of the 3[rd] nonlinear harmonic radiation is about 1% of the fundamental. Thus, novel schemes [14-17] have been proposed for enhancing the harmonic efficiency in high-gain FEL, some of which are being considered for experimental demonstration.

    Recently, the low-gain oscillator configuration has been reconsidered as a promising candidate for a hard x-ray FEL [18] through the use of a high-reflectivity high-resolution x-ray crystal [19] cavity, combined with ultra-low emittance electron beams from an energy-recovery linac (ERL). With the peak brilliance comparable to SASE and the average brilliance several orders of magnitude higher than SASE, the x-ray FEL oscillator (XFELO) in the range of 10-25keV may open up new scientific opportunities in various research fields, such as x-ray imaging and especially angle-resolved photo-emission spectroscopy based time-resolved measurement of Fermi surfaces. Harmonic lasing in the low-gain oscillator was predicted in 1980 [20]. The first experimental demonstrations of 3[rd] harmonic lasing were obtained using infrared FEL oscillators in 1988 [21, 22]. Subsequently, infrared generation on the 2[nd] [23], 3[rd] and 5[th] [24] harmonic lasing were demonstrated at Jefferson Lab. More recently, the Novosibirsk terahertz FEL successfully lased on the 3[rd] harmonic [25]. In this letter, we propose a harmonic lasing scheme of XFELO and discuss the possibility to generate hard x-ray FEL with an intermediate energy electron beam. The key point is that the Bragg energy of the crystal mirrors is selected for the photon energy of the undulator harmonic radiation, rather than the fundamental. Studies show that

coherent FEL radiation with megawatt (MW) peak power, in the spectral region of 10-25keV, can be generated with a 3.5GeV electron beam from an ERL or a high repetition superconductive linac.

The challenges of harmonic lasing in an XFELO are, firstly, a single-pass harmonic gain higher than the cavity losses which is determined by the electron beam quality, and, secondly, a more favorable lasing relation for the high harmonic than for the fundamental one. Generally, switching between the lasing regimes on fundamental or high harmonics can be obtained by a simple manipulation with the optical cavity. As shown in Fig. 1, the Bragg energy $E_H$ of the crystal mirrors in XFELO is set to the photon energy of the interested harmonic, instead of that of the fundamental radiation. Harmonic lasing of XFELO could then be realized because of the high selectivity of the x-ray crystal mirrors. Moreover, as illustrated in Fig. 2, if a four-crystal cavity is used, it will hold a promising wavelength tuning ability for a harmonic lasing XFELO [26].

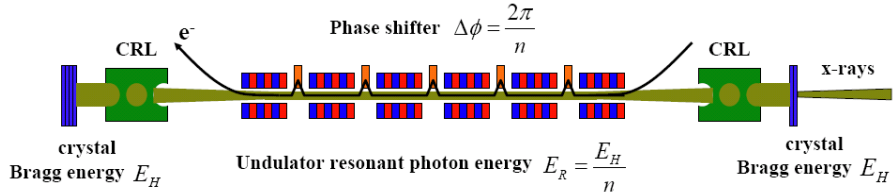

FIG. 1. Harmonic lasing scheme of an x-ray FEL oscillator.

Since XFELO requires several segments of undulator to obtain a sufficient single-pass gain, a generic phase shifter is set for fine tuning of phase matching between undulator segments. In reality, FEL performance in the proposed harmonic lasing XFELO depends on the ratio of the mirror losses for the fundamental and harmonic, the beam parameters and the undulator length. If the fundamental radiation still has large net gain after being filtered by the crystal, a method to suppress exponential growth of the fundamental [16] can be used. In this method, the value of the phase jump, $\Delta\varphi$, applied in each phase shifter is set to $2\pi/n$, such that the electron interaction with the fundamental is disrupted, while the $n^{th}$ harmonic interaction evolves unhindered

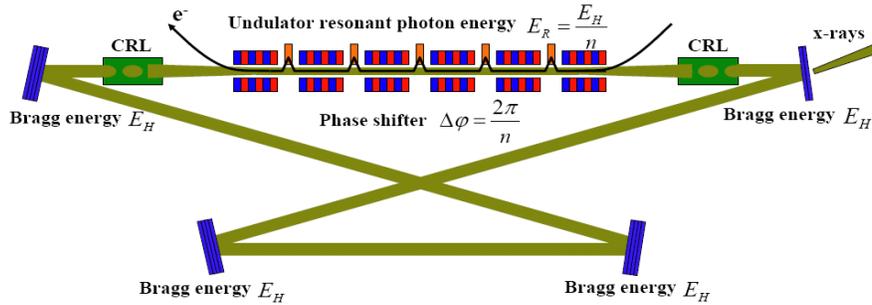

FIG. 2. Harmonic lasing scheme of a four-crystal x-ray optical cavity allowing a broad range of wavelength tuning.

To obtain harmonic lasing in an XFELO, the first limiting factor is to generate a high brightness beam with sufficient single-pass gain at harmonic wavelengths. As shown in Ref. 18, the high coherence mode of an ERL and the low charge mode of a high repetition superconductive linac are well suited for an XFELO. The electron beam parameters assumed here are: bunch charge of 20pC, normalized transverse emittance of 0.083μm-rad, peak current less than 100A, and slice energy spread of 100keV. Since these beam parameters are invariant under acceleration, we assume a 3.5GeV superconductive linac with a micro-pulse repetition rate of 1MHz, similar to the linac of FLASH [27] and the European x-ray FEL [4]. The 20pC charge operation [28] has already been demonstrated in LCLS with a peak current of several kA, and a similar low charge operation has also been investigated numerically for the European x-ray FEL. We are optimistic about achieving the above beam parameters since the peak current in our case is two orders of magnitude lower. Further "start-to-end" simulation studies, based on the beam dynamics of the European x-ray FEL, support the beam parameters used here.

TABLE 1. The main parameters of harmonic lasing XFELO.

| Parameters | 3$^{rd}$ harmonic | 5$^{th}$ harmonic |
|---|---|---|
| Crystal Bragg energy $E_H$ | 12.42keV | 20.71keV |
| Phase jump $\Delta\varphi$ | $4\pi/3$ | $6\pi/5$ |
| Undulator period $\lambda_u$ | 15mm | 15mm |
| Undulator number $N_u$ | 1200 | 1200 |
| Undulator parameter $K$ | 1.3244 | 1.3244 |
| Beam energy $E$ | 3.5GeV | 3.5GeV |
| Slice energy spread $\sigma$ | 100keV | 100keV |
| Beam peak current $I$ | 20A | 100A |
| Slice emittance $\varepsilon_n$ | 0.083μm-rad | 0.083μm-rad |
| Single-pass gain $g_h$ | 65% | 72% |
| Total Cavity reflection $r$ | 80% | 80% |
| Cavity length $L_c$ | 150m | 150m |
| Bragg crystal | C(4,4,4) | C(5,5,9) |
| FWHM spectral width | 5.5meV | 24.6meV |
| FWHM temporal width | 463fs | 107fs |
| Photons / pulse | $0.86\times10^8$ | $0.24\times10^8$ |
| Output peak power | 0.35MW | 0.74MW |

Table 1 gives two examples of XFELO with layout of Fig. 1, corresponding to the 3$^{rd}$ and the 5$^{th}$ harmonic lasing. The beam parameters are assumed to be those listed above. The undulator parameters are chosen to be resonant at a fundamental wavelength of 0.3nm. Thus the 3$^{rd}$ and the 5$^{th}$ harmonics of the undulator radiation are 0.1nm (12.42keV) and 0.06nm (20.71keV), respectively. Fig. 3 illustrates the numerically simulated single-pass gain of the harmonics, with respect to different phase jump between undulator segments, where the parameters for 5$^{th}$ harmonic lasing from Table 1 were used. For simplification, optimal phase jump of $4\pi/3$ and $6\pi/5$ were set for the 3$^{rd}$ and 5$^{th}$ harmonic lasing of XFELO, such that single-pass gain of the interested harmonic is maximized while those of other harmonics are successfully suppressed.

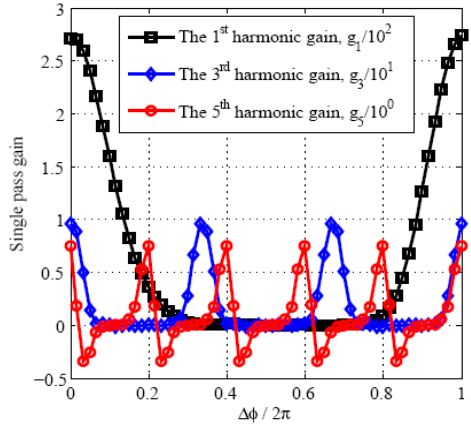

FIG. 3. The single-pass gain of the 1$^{st}$, 3$^{rd}$ and 5$^{th}$ harmonic .vs. different phase jump between undulator segments.

To trap the radiation and maximize the FEL gain, the harmonic lasing of XFELO must provide appropriate transverse focusing, effective reflection and transmission output coupling of the interested harmonic radiation. This can be accomplished with presently available compound refractive lenses (CRL) and Bragg crystals with appropriate thickness, as shown in Figs. 1 and 2. The total reflectivity and output coupling of the XFELO cavity are assumed to be 80% and 5%, respectively. Since $(1+g_h)\times r>1$, where $g_h$ is the single-pass gain of the interested harmonic and $r$ is the round-trip cavity reflectivity, the harmonic radiation evolves from initial spontaneous emission to a coherent pulse. After an exponential growth, the harmonic intensity saturates when $(1+g_h)\times r=1$ because of the gain decrease caused by over-modulation due to the high intensity intra-cavity radiation. In the following discussion, only the results of the 3$^{rd}$ harmonic lasing case in Table 1 are given and discussed for simplification.

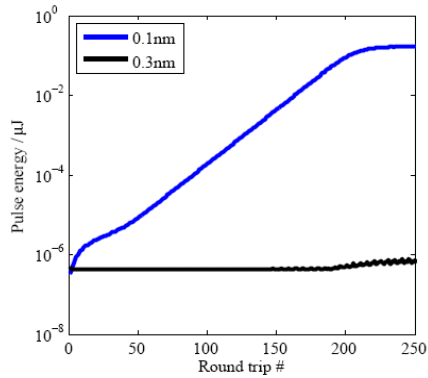

FIG. 4. The 1$^{st}$ and 3$^{rd}$ harmonic radiation energy growth in the 3$^{rd}$ harmonic lasing case of XFELO.

To study the harmonic lasing of XFELO, GENESIS [29] was modified to enable import of an external harmonic field distribution. The propagation of the x-ray radiation in free space between the undulator and the crystal, the reflection by the Bragg mirrors and focusing are modeled using OPC [30]. The frequency filtering of the Bragg crystal is carried out in the spectral domain and transformed back to the time domain. Under such circumstances, one full time-dependent tracking from initial spontaneous emission to final saturation takes about 3 months for 1 CPU. Fig. 4 shows the radiation pulse energy as a function of the pass number. An exponential growth of the 3rd harmonic emerges from the initial shot noise start-up after about 50 passes, while the fundamental does not grow significantly.

A typical example of the 12.42keV radiation evolution from initial shot noise to saturation is illustrated in Fig. 5. The first row of panels displays the temporal power profile of the output pulse after various pass numbers, while the second row shows the corresponding spectrum. According to the simulation, 12.42keV radiation with 0.35MW peak power can be generated in the 3rd harmonic lasing of XFELO, with the parameters listed in Table 1. The total energy per pulse coupled out of the cavity is 0.17μJ, corresponding to $0.86\times10^8$ photons. Fig. 5 demonstrates a temporal width of 463fs FWHM and a bandwidth of 5.5meV FWHM. This corresponds to a time-bandwidth product of 0.62, which is close to the Fourier transform limit of 0.44 for a Gaussian pulse profile.

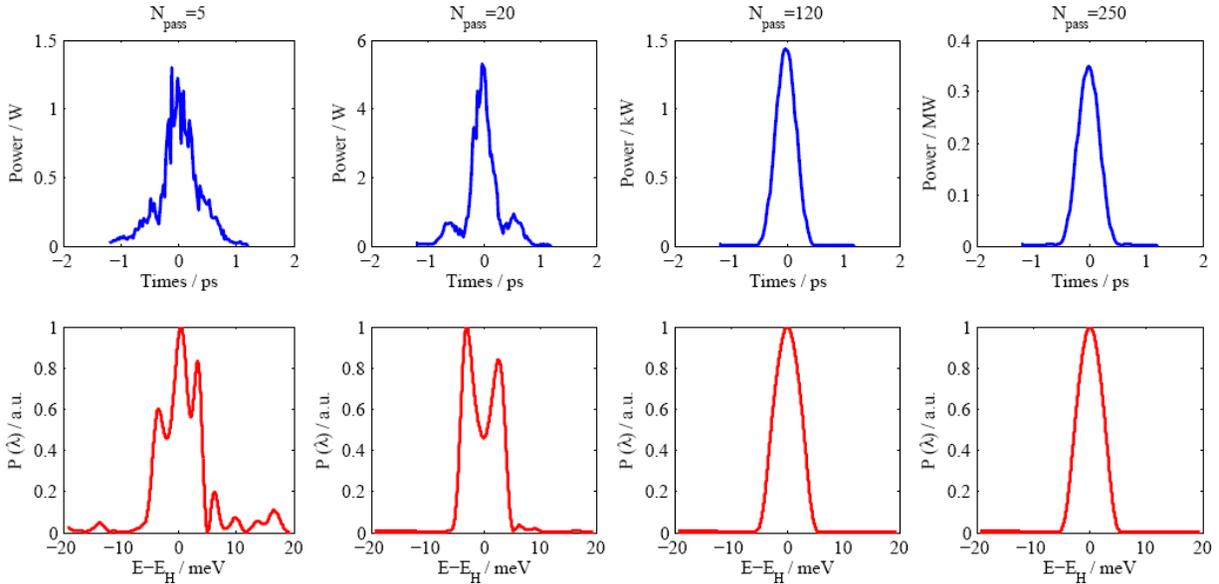

FIG. 5. The temporal and spectral profile of the 3rd harmonic in the 3rd harmonic lasing case of XFELO.

The robustness of harmonic lasing of an XFELO is compared with the normal XFELO. For a particular wavelength, the requirements on the cavity tolerance, e.g. an angular tilt less than 50nrad, in both schemes are similar in principle and in steady-state simulation. The sensitivity on the electron beam parameters in the harmonic lasing scheme is approximately 3 times more serious than in normal XFELO, in order to obtain a sufficient single-pass gain at the higher harmonic. The gain reduction induced by the beam imperfection could be compensated by using a short Rayleigh range cavity, which however may degrade the XFELO cavity stability. A lower x-ray saturation power is expected in the XFELO cavity for harmonic lasing, corresponding to 3.4μJ pulse energy and 3.4W average power in our case. Therefore, the thermal load effects on the crystal will be much more relaxed, and a linear thermal expansion coefficient less than $1\times10^{-7}$ K$^{-1}$ is sufficient to ensure a stable x-ray cavity. In addition, the induced electron beam energy spread from the FEL interaction is several times smaller in the harmonic lasing XFELO, thus mitigating the beam loss in the ERL return arc and making ERL operation more straightforward.

In summary, compared with high gain SASE and low-gain XFELO operated at the fundamental frequency, harmonic lasing of an XFELO is able to produce significant 10-25keV x-ray FEL radiation using an intermediate energy beam of 3.5GeV. For the 3rd harmonic lasing of XFELO operating in the hard x-ray spectral region, the numerical example in this letter demonstrates a peak brilliance of $1.79\times10^{32}$ photons/(s mm$^2$ mrad$^2$ 0.1%BW) which is comparable with that of a high gain SASE, e.g. $8.5\times10^{32}$ photons/(s mm$^2$ mrad$^2$ 0.1%BW) for LCLS [2], and an average brilliance 3 orders of magnitude higher than SASE. With the predicted characteristics, harmonic lasing of an XFELO would contribute to the simplification of x-ray FELs, and may open up new scientific opportunities in various research fields.

The authors would like to thank M. Zhang and Q. Gu for beam dynamic simulations, B. Liu and P. J. M. van der Slot for their helps in OPC implement, L. P. Chen for thermal loading issue, J. H. Wu for enthusiastic discussions on FEL brilliance, and D. J. Dunning for useful discussions and for improving the writing. This work is supported by Natural Science Foundation of China under Grant No. 11175240 and Shanghai Natural Science Foundation under Grant No. 09JC1416900.

*Corresponding author: denghaixiao@sinap.ac.cn